
\magnification=1200


\hsize 16truecm
\vsize 24truecm
\def\build#1_#2^#3{\mathrel{\mathop{\kern 0pt#1}\limits_{#2}^{#3}}}
\font\tenll=lasy7
\newfam\llfam
\textfont\llfam=\tenll
\def\ll{\fam\llfam\tenll}
\def\carre{{\ll 2}}
\def\n{\noindent}
\def\m{\medskip}

\font\fiveeu=eufm5
\newfam\eufam
\textfont\eufam=\fiveeu

\def\am{a_{\mu}}

\def\n{\noindent}
\def\m{\medskip}

\hsize 17truecm
\vsize 24truecm
\font\twelve=cmbx10 at 13pt
\baselineskip 18pt
\nopagenumbers

\null
\vskip 5truecm

\centerline{\twelve HADRONIC CONTRIBUTIONS TO THE MUON $g-2$}
\centerline{\twelve AND LOW-ENERGY QCD}

\centerline{Eduardo de RAFAEL}

\vskip 4truecm
{\leftskip=1cm
\rightskip=1cm
\centerline{\bf Abstract}

\medskip

The contributions to the muon anomalous magnetic moment from hadronic
vacuum polarization and from hadronic light-by-light scattering are
reexamined within the frame work of chiral perturbation theory; the
$1/N_c$-expansion; and the extended Nambu Jona-Lasinio model of
low-energy QCD.
\par}

\vskip 6truecm

\noindent October 1993

\noindent

\vfill\eject



\pageno=1

The anomalous magnetic moment of the muon is known to an accuracy of
less than $10$ppm. The latest result from the combined $\mu^+$ and
$\mu^-$ measurements performed at CERN is [1]
$$a_{\mu}\equiv{1\over 2}
(g_{\mu}-2) = 11\ 659\ 230(85)\times 10^{-10}.\eqno (1)$$

\noindent The standard model prediction quoted in the most recent
review article on the subject [2] is
$$a_{\mu}^{SM}=11\ 659\
192.8(17.6)\times 10^{-10}.\eqno (2)$$

\noindent This includes the pure QED contributions from leptons to
four-loops in perturbation theory, as well as the lowest order weak
interaction contribution
$$a_{\mu}^{weak}=  19.5(1.0)\times
10^{-10};\eqno (3)$$
and the contributions from hadronic vacuum
polarization  (Figs. 1a and 1b), and from the hadronic
light-by-light scattering (Fig. 2). These hadronic contributions
amount to a total
$$a^H_{\mu}= 702.7(17.5)\times 10^{-10}.\eqno (4)$$

\noindent The agreement between the standard model prediction and the
present experimental value is rather good.

A new high-precision experiment to measure $a_{\mu}$ with an
expected error of $\pm 4\times 10^{-10}$ is underway at BNL.
To confront usefully theory with the new experiment will require a
better determination of the hadronic contributions. A substantial
improvement in the accuracy of the contribution from the hadronic
vacuum polarization is possible. It requires, essentially, a more
precise determination of the low energy tail of the total $e^+e^-\to$
hadrons cross-section. The VEPP-2M facility currently in operation,
and DA$\Phi$NE in the near future, can attain that precision. By
contrast, the contribution to $a_{\mu}$ from the hadronic
light-by-light scattering, which  we shall denote $a_{\mu}(had \ 2)$,
cannot be expressed as a convolution of experimentally accessible
observables. Here, we have to resort to theoretical wisdom The result
quoted in [2], based on the estimates made by Kinoshita,
Ni$\breve{\hbox{z}}$i\'c and Okamoto (KNO) [3], is
$$a_{\mu}(had \ 2)=4.9(.5)\times 10^{-10}\ .\eqno (5)$$

Doubts on the reliability of the accuracy of the
hadronic light-by-light scattering contribution above have been
recently raised in refs. [4] and [5]. As emphasized by Martin
Einhorn [5], {\it the issue is important for the interpretation of
the new experiment and crucial for drawing inferences about potential
new physics}. The purpose of this note is to analyze the hadronic
contributions to $a_{\mu}$ from the point of view of low energy QCD
and to show that, within this framework, it might be possible to
improve the determination of $a_{\mu}(had\ 2)$ to a sufficiently
safe accuracy for a useful interpretation of the new BNL-experiment.

\bigskip

\n {\bf 1.}\hskip 0,5truecm I suggest we look first at the well
known contribution from hadronic vacuum polarization in Fig. 1a which
we shall denote $a_{\mu}(had\ 1a)$. It is usually expressed in the
form of an integral representation $(e^2=4\pi\alpha,\hbar=c=1)$
$$a_{\mu}(had\
1a)={\alpha\over\pi}\int^{\infty}_{0}{dt\over t}
K(t/m^2_{\mu})e^2{1\over\pi}Im\Pi^{(H)}(t)\ ,\eqno (6)$$
which is a
convolution of the hadronic spectral function $\displaystyle {1\over
\pi}Im\Pi^{(H)}(t)$, related to the total $e^+e^-\to$ hadrons
cross-section $\sigma(t)\ (m_e\to 0)$ by
$$\sigma(t)={4\pi^2\alpha\over t}e^2{1\over \pi}Im\Pi^{(H)}(t)\
,\eqno (7)$$
with the QED function
$$K(t/m^2_{\mu})=\int^{1}_{0}dx{x^2(1-x)\over x^2+(1-x)t/m^2_{\mu}}\
,\eqno (8)$$
which is positive and monotonically decreasing from
$t=0$ to $t=\infty$. I propose we try to calculate $a_{\mu}(had\
1a)$, {\it without using the experimentally known cross-section}
$\sigma(t)$. This calculation will then serve us as a test of how
well we understand - theoretically - $a_{\mu}(had\ 1a)$. (If here we
fail to reproduce the phenomenological determination, there is
little hope to expect that we can make a reliable estimate of
$a_{\mu}(had\ 2)$.)

For our purposes, it is also convenient to express $a_{\mu}(had\ 1a)$
using the integral representation [6]
$$a_{\mu}(had\ 1a)={\alpha\over \pi}\int^{1}_{0}dx (1-x)\left[-e^2
\Pi^{(H)}_R\left({x^2\over 1-x}m^2_{\mu}\right)\right]\ ,\eqno (9)$$
where $\Pi^{(H)}_R(Q^2)$ denotes the renormalized
$(\Pi^{(H)}_R(0)=0)$  hadronic photon self-energy. The link between
the two integral representations in (6) and (9) is provided by the
once subtracted dispersion relation $(Q^2\equiv -q^2\geq 0$ in our
metric $+,-,-,-)$
$$\Pi^{(H)}_R(Q^2)=\int^{\infty}_{0}{dt\over
t}{q^2\over t-q^2}{1\over\pi}Im\Pi^{(H)}(t)\ .\eqno (10)$$
The bulk
of the integral in (9) is governed by the low energy behaviour of
$\Pi^{(H)}_R(Q^2)$. The typical momemtum of the virtual photon in
Fig. 1a is $Q^2\sim m^2_{\mu}$. These are momenta values much
smaller than the characteristic scale $\Lambda_{\chi}$ of the
spontaneous chiral symmetry breaking in QCD with three light
flavours $(\Lambda_{\chi}\simeq 1 GeV)$. Therefore, the appropriate
way to look at this problem is within the framework of the low energy
effective field theory of QCD i.e., chiral perturbation theory
($\chi$PT).

Since $\Pi^{(H)}_R(0)\equiv 0$, the leading contribution to
$\am(had\ 1a)$ from tree level couplings in $\chi$PT comes from the
$O(p^6)$ effective Lagrangian ; and more precisely from the term
proportional to $F_{\mu\nu}\carre F^{\mu\nu}$. The coupling constant
of this term is related to the slope at the origin of the hadronic
photon self-energy. The relevant effective Lagrangian we are
concerned with here is
$${\cal L}_{eff}=-{1\over
4}\left\{F_{\mu\nu}F^{\mu\nu}-P_1e^2\partial^{\lambda}F^{\mu\nu}
\partial_{\lambda}F_{\mu\nu}+...\right\}\eqno (11)$$
where
$$\left.P_1=-{\partial\Pi^{(H)}_R(Q^2)\over\partial
Q^2}\right\vert_{Q^2=0}\ .\eqno (12)$$
In QCD, and in the large
$N_c$ (number of colours) approximation [7], the constant $P_1$
adquires a leading contribution of $O(N_c)$. We can then conclude
that to leading order in the $1/N_c$-expansion of QCD and in the
chiral limit, to lowest non-trivial order in $\chi$PT, the
calculation of $a_{\mu}(had\ 1a)$ can be reduced to the
determination of only one coupling constant of the $O(p^6)$ chiral
effective Lagrangian of QCD, with the result [8]
$$a_{\mu}(had\
1a)\simeq\left({\alpha\over \pi}\right)^2m^2_{\mu} {4\pi^2\over
3}P_1\ .\eqno (13)$$

What do we know about the constant $P_1$? As already shown, $P_1$ is
an $O(p^6)$ coupling and, unfortunately, our present knowledge in
$\chi$PT stops at $O(p^4)$. The coupling constants of the $O(p^4)$
effective chiral Lagrangian of QCD have been determined
phenomenologically in a series of beautiful papers by Gasser and
Leutwyler [9]. It has subsequently been shown [10, 11] that these
constants are practically saturated by the lowest  resonance
exchanges and particularly by vector-exchange whenever vectors can
contribute. We can adopt this phenomenological observation to obtain
an estimate of $P_1$. Using the most natural description of vector
mesons in terms of vector fields as discussed e.g. in ref. [12], one
obtains the result
$$P_1={4\over 3} {f_{V}^2\over M_{V}^2}\ ,\eqno
(14)$$
where $f_V$ denotes the coupling of vector mesons to the
electromagnetic field $(f_V\simeq 0.2)$ and $M_V$ is the vector
meson mass $(M_V\simeq 770 MeV)$. The resulting estimate of
$a_{\mu}(had\ 1a)$ in eq. (13) is then
$$a_{\mu}[had\ 1a, \hbox{ eqs.
(13, 14)}]\simeq 7.1\times 10^{-8}\ ,\eqno (15)$$
to be compared to
the latest phenomenological determinations [3, 13, 14]
$$a_{\mu}(had\ 1a)=\left\{\normalbaselineskip=18pt\matrix{
7.068 (0.059)(0.164)\times 10^{-8}\ ,&\hfill(a)\cr
6.84 (0.11)\times 10^{-8}\ ,\hfill &\hfill(b)\cr
7.100 (0.105)(0.49)\times 10^{-8}\hfill &\hfill(c)
}\right.\eqno(16)$$
where in (16a) and (16c) the
first error is statistical and the second systematic. The agreement
of our $\chi$PT estimate with the full phenomenological determination
is embarrassingly good.

At this point some readers will object that the estimate in (15)
uses, after all, phenomenological input - the ratio $f^2_V/M^2_V$. I
want to discuss next, as a possible alternative, how well the
extended Nambu Jona-Lasinio model (ENJL-model) of low energy QCD,
which has recently been developped in refs. [15] and [16], does in
calculating $a_{\mu}(had\ 1a)$. The ENJL-model can be viewed as an
approximation of large-$N_c$ QCD, where the only new interaction
terms retained  after integration of the high frequency modes of the
quark and gluon fields down to a scale $\Lambda_{\chi}$ at which
spontaneous chiral symmetry breaking occurs, are those which can be
cast in the form of four-fermion ope\-ra\-tors. The parameters of the
model are then $\Lambda_{\chi}$ and the two coupling constants $G_s$
and $G_v$ of the two-possible (scalar-pseudoscalar) and
(vector-axial) four-fermion cou\-plings. These two couplings can be
traded for the mass $M_Q$ of the constituent chiral quark, which
appears as a non-trivial solution to the gap equation involving
$G_s$; and the effective\break axial coupling $g_A$ of the
constituent chiral quarks to the pseudoscalar Goldstone bosons [17]
$$g_A={1\over
1+4G_v{M^2_Q\over\Lambda^2_{\chi}}\Gamma(0,{M^2_Q\over
\Lambda^2_{\chi}})}\ ,\eqno (17)$$
where $\Gamma(0,\varepsilon)$
denotes the incomplete gamma function
$$\Gamma(n,\varepsilon)=\int^{\infty}_{\varepsilon}{dz\over z}e^{-z}
z^n\ .\eqno (18)$$

In ref. [16] the QCD vector two-point function, within the
approximation provided by the ENJL-model, has been calculated to all
orders in powers of momenta at the leading order in the
$1/N_c$-expansion with the result
$$\Pi^{(1)}_V={\overline{\Pi}^{(1)}_V(Q^2)\over1+Q^2{8\pi^2 G_v\over
N_c\Lambda^2_{\chi}} \overline{\Pi}^{(1)}_V(Q^2)}\ ,\eqno (19)$$
where
$$\overline{\Pi}^{(1)}_V(Q^2)={N_c\over
16\pi^2}8\int^{1}_{0}dy\ y(1-y)\Gamma\left(0,{M^2_Q+Q^2y(1-y)\over
\Lambda^2_{\chi}}\right)\ .\eqno (20)$$
In terms of $\Pi^{(1)}_V$,
the hadronic photon self-energy is given by the expression $(\Sigma
Q^2_i=2/3, i=u,d,s)$
$$\Pi^{(H)}_R(Q^2)={2\over
3}\left(\Pi^{(1)}_V(Q^2)-\Pi^{(1)}_V(0)\right)\ ;\eqno (21)$$
and the
$O(p^6)$ coupling constant $P_1$ defined in eqs. (11) and (12) can
then be easily obtained:
$$P_1(\hbox{ENJL})={N_c\over 16\pi^2}{2\over
3}{1\over M^2_Q}{4\over
15}\left[\Gamma\left(1,{M^2_Q\over\Lambda^2_{\pi}}\right)+{5\over
4}{1-g_A\over g_A}\right]\ .\eqno (22)$$
In the limit $G_v\Rightarrow
0$ (i.e., $g_A=1$) and $\Lambda^2_{\chi}>> M^2_Q$ we recover the
result of the  constituent chiral quark model [18], [19]. Using the
input parameter values
$$M_Q=265 MeV,\ \Lambda_{\chi}=1165 MeV,\
g_A=0.61\eqno (23)$$
corresponding to the fit 1 in ref. [15] one gets
$$\am\left[had\ 1a\ ; \hbox{ eqs. (13, 22)}\right]=6.7\times
10^{-8}\eqno (24)$$
while, using the same $M_Q=265 MeV$ value, the
constituent quark model prediction comes out to be rather low $(3.9
\times 10^{-8})$. We conclude that the ENJL-model does well in
reproducing the semiphenomenological estimate of eq. (15).

The deep reason why I am discussing the ENJL-model is that, as I
shall next show, it offers the possibility to go beyond the leading
$O(p^6)$ contribution of $\chi$PT. When, later, we discuss the
hadronic light-by-light contribution, this will appear to be a
crucial issue.

In trying to estimate the $O(p^8)$ correction from the hadronic
photon self-energy to the integral in eq. (9) there appears an
interesting difficulty. Once we factorize the second derivative at
the origin of the photon self-energy in eq. (9), the remaining
integral over the $x$-variable diverges logarithmically in the
ultraviolet region $(x\to 1)$. In the language of $\chi$PT this
means that a genuine local counterterm of the type
$\overline{\psi}\sigma^{\mu\nu}\psi F_{\mu\nu}$ has to be introduced,
with a constant which after renormalization will absorb the
$UV$-divergence of the previous loop integral. However, the finite
part of this new constant is not fixed by arguments of symmetry and
depends on the details of the dynamics of the underlying theory. It
is precisely in this sense that the ENJL-model helps, {\it without
appealing to extra phenomenological input}. If we insert the full
expression for the hadronic photon self-energy given by eqs.(21, 19,
20) in the integrand of eq.(9), the integral over the $x$-variable is
convergent. Strictly speaking, the $Q^2$-dependence of the hadronic
photon self-energy predicted by the ENJL-model is expected to be
adequate up to $Q^2$ values smaller than $\Lambda^2_{\chi}$. Beyond
that we enter a regime where a description in terms of perturbative
QCD is certainly more appropriate. We have here another observable
which, like the $\pi^+-\pi^0$ electromagnetic mass difference
recently discussed in ref. [16] (see also references therein),
offers the possibility to test the quality of the matching between
long-distance behaviour and short-distance behaviour. I don't want
to discuss this in detail now, because it would take us too far away
from the main purpose of this note. Here, I shall only give the
result I get for the total $\am(had\ 1a)$ using the ENJL-model
expression of $\Pi^{(H)}_R(Q^2)$ for the evaluation of the
long-distance part contribution; and using perturbative QCD for the
calculation of the short-distance part contribution. The matching
between short-and long-distances is defined by the optimal choice of
an $\widehat x$ in the integral (9) which minimizes the variation of
the total long $\displaystyle(\int^{\widehat x}_{0})$ - plus short
$\displaystyle(\int^{1}_{\widehat x})$ - distance contribution. With
the same input parameter values as in eq.(23), I find
$$\am(had\ 1a, \hbox{ ENJL })=6.7\times 10^{-8}\ .\eqno (25)$$
This is
the result corresponding to the optimal choice $\widehat x\simeq
0.9$. The stability around this value is very good. (Perhaps I should
clarify that, in terms of the $t$-variable in the hadronic spectral
function representation of eqs.(6, 7, 8), the value $\widehat x\simeq
0.9$ corresponds to an ''equivalent'' $\widehat t\simeq 2 GeV^2$;
which is very reasonable). The fact that the two results in (24) and
(25) are exactly the same is a numerical coincidence. If instead of
using the values of the input parameters in (23), I use the values
corresponding to fit 2 in [15] (i.e., $M_Q=263 MeV,\
\Lambda_{\chi}=1048 MeV,\ g_A=0.62$) the result is $\am(had\ 1a,
\hbox{ENJL fit }2)=6.6\times 10^{-8}$.

So far, I have not discussed the explicit contributions to
$\am(had\ 1a)$ from $\pi^+\pi^-$ (and $K^+K^-$) vacuum polarization
insertions in Fig.1a. In the framework of $\chi$PT, these
contributions appear as chiral loops induced by the lowest order
effective Lagrangian
$${\cal L}_{eff}={1\over 4} f^2_{\pi} tr
D_{\mu}UD^{\mu}U^{\dagger}+...\ ,\eqno (26)$$
where $U$ is a unitary
$3\times 3$ matrix with det $U=1$ which collects the pseudoscalar
fields and $D_{\mu}U$ the covariant derivative
$$D_{\mu}U=\partial_{\mu}U-i(v_{\mu}+a_{\mu})U+iU(v_{\mu}-a_{\mu})
\eqno (27)$$
with $v_{\mu}(a_{\mu})$ external vector (axial) field
sources. In our case : $v_{\mu}=eQA_{\mu}$ and $a_{\mu}=0$, where
$Q=1/3$ diag $(2,-1,-1)$ and $A_{\mu}$ is the electromagnetic
gauge-field. The Lagrangian in (26) has an effective coupling
$$-ie\
A_{\mu}\left(\pi^+\build {\partial_{\mu}}_{}^{\leftrightarrow}\pi^-+
K^+\build {\partial_{\mu}}_{}^{\leftrightarrow}K^-\right)\ ,\eqno
(28)$$
which leads to the lowest order (i.e., $O(p^4)$) hadronic
spectral function in $\chi$PT :
$${1\over \pi}Im\Pi^{(H)}_{\chi
loop}(t)={1\over 16\pi^2}{1\over 3}\left(1-{4m^2_{\pi}\over
t}\right)^{3/2}\Theta(t-4m^2_{\pi})+\pi\leftrightarrow K\ .\eqno
(29)$$
There are two important features emerging here. First, the
fact that, by contrast to the previous contributions we have
discussed, this spectral function is non-leading in the
$1/N_c$-expansion. (This is in fact a rather general property of
chiral loops in QCD.) Second, the observation that when ${1\over
\pi}Im\Pi^{(H)}_{\chi loop}$ is inserted in the integral
representation in eq.(6), although nominally it is a leading
$O(p^4)$ term in chiral power counting, it contributes a very small
amount to $a_{\mu}(had\ 1a)$. Numerically, I find
$(\pi^+\pi^-+K^+K^-)$
$$a_{\mu}(had\ 1;\chi loops)=(0.71+0.07)\times
10^{-8}\ .\eqno (30)$$
This small result is essentially due to the
empirical fact that the physical threshold for $\pi^+\pi^-$ (and
$K^+K^-$) production is much bigger than the muon mass, therefore
suppressing further the, already $1/N_c$-suppressed, $O(p^4)$ chiral
loop contribution. More quantitatively, for $t\geq 4m^2_{\pi}$ the
function $K(t)$ in eq.(8) is very well approximated by its
assymptotic ${1\over 3}m^2_{\mu}/t$ - behaviour. Therefore, the
contribution from chiral loops is practically dominated by the slope
of their contribution to the photon self-energy. The slope in
question is
$$P_1(\chi loop) ={1\over 16\pi^2}{1\over 30}\left({1\over
m^2_{\pi}}+{1\over M^2_k}\right)\ ,\eqno (31)$$
about one order of
magnitude smaller than the leading $O(N_c)$ slope values in eqs.(14)
and (22). The result in eq.(30) is the leading chiral loop
correction, and as such has to be  added to our full estimate in the
chiral limit of the $O(N_c)$ contribution in eq.(25). This sum gives
the best theoretical estimate I can think of at present for
$a_{\mu}(had\ 1a)$ - {\it without using the empirical knowledge of
the $e^+e^-\to$ hadrons total cross-section - i.e.,}
$$a_{\mu}(had\
1a, \hbox{ Theory })=7.5\times 10^{-8}\ .\eqno (32)$$
It is in the
high ball park of the phenomenological determinations in (16$a,\ b,\
c$), but not bad at all. We conclude that we understand reasonably
well, within the framework of low energy QCD, $a_{\mu}(had\ 1a)$.

\bigskip

\n {\bf 2.}\hskip 0,5truecm It is time to examine now the
contribution from the hadronic light-by-light scattering in Fig.2
i.e. $a_{\mu}(had\ 2)$. Again, the typical momenta of the virtual
photons in Fig.2 is $Q^2\sim m^2_{\mu}$; and the appropriate
framework to analyze the problem is the one of $\chi$PT. At first
sight, the leading contribution to $a_{\mu}(had\ 2)$ from direct tree
level couplings appears to come from the $O(p^8)$ chiral effective
Lagrangian with four electromagnetic field strength tensors i.e., the
Lagrangian $\left(\Sigma Q^4_i=2/9,\quad i=u,\ d,\ s\right)$
$$\eqalign{{\cal L}_{eff}=&{N_c\over 16\pi^2}{1\over 360}{1\over
M^4_Q}{2\over
9}e^4\left\{4h_1(F_{\mu\nu}(x)F^{\mu\nu}(x))^2\right.\cr
&\left.+7h_2(\varepsilon^{\mu\nu\rho\sigma}F_{\mu\nu}(x)
F_{\rho\sigma}(x))^2\right\}\cdot\cr}\eqno(33)$$
\noindent This is the QCD-analogue of the famous
Euler-Heisenberg effective Lagrangian of QED. [20]. (For  an
excellent book on the subject see ref. [21].) With the normalization
I have chosen, the coupling constants $h_1$ and $h_2$ are
dimensionless. In the constituent chiral quark model [18,19,15]
$h_1=h_2=1$. In QCD these effective couplings appear to leading
$O(N_c)$ in the $1/N_c$-expansion. (This is why I have already pull
out the $N_c$-factor in (33).) The direct insertion in the muon
vertex diagram of Fig. 2 of the local light-by-light scattering
couplings given by the effective Lagrangian in (33) leads however to
$UV$-divergent integrals. The situation here is rather similar to
the one we already encountered in the case of hadronic vacuum
polarization when one tries to insert the local $O(p^8)$ photon
self-energy coupling. Again, what we find now is that we need more
information about the underlying QCD dynamics than just the two
local couplings $h_1$ and $h_2$. In view of its phenomenological
successes it seems reasonable to expect that the ENJL-model we
discussed earlier may provide the required help. More on that later.
First, let us examine other possible contributions to $a_{\mu}(had\
2)$ within the framework of $\chi$PT.

There is of course the contribution from the hadronic light-by-light
scattering amplitude induced by the effective lowest order coupling
in (26). Like in the case of the hadronic vacuum polarization, it
gives rise to an $O(p^4)$ contribution, which is also non-leading in
the $1/N_c$-expansion. The chiral loop calculation of the
light-by-light scattering tensor
$\Pi^{\mu\alpha\beta\gamma}(k,q_1,q_2,q_3)$ leads to a rather
complicated expression. In order to get some feeling, we can however
consider the limit where all the photon momenta are negligible with
respect to the physical pion mass. This limit is described by the
Euler-Heisenberg effective Lagrangian of scalar-QED [21] i.e.,
$$\eqalignno{&{\cal L}_{eff}\hbox{(scalar-QED) }=-{1\over 4}
F_{\mu\nu}F^{\mu\nu}+{1\over 16\pi^2}{1\over 30}{1\over
m_{\pi}^2}e^2\partial^{\lambda}F^{\mu\nu}\partial_{\lambda}F_{\mu\nu}
\cr
&+{1\over 16\pi^2}{1\over 1440}{1\over
m_{\pi}^4}e^4\left\{7(F_{\mu\nu}F^{\mu\nu})^2+
(\varepsilon^{\mu\nu\rho\sigma}F_{\mu\nu}F_{\rho\sigma})^2\right\}
+...&(34)\cr}$$
As already discussed earlier, the second term in the
r.h.s. above governs the slope of the photon self-energy induced by
the lowest order chiral loop (see eq.(31)). The third term is the
resulting one for the effective light-by-light scattering. From the
comparison between (33) and (34) one can see that for values
$h_1\simeq h_2\simeq O(1)$, the order of magnitude of the direct
$O(N_c)$ couplings are about the same size (or smaller for the
$(FF)^2$ - term) than those induced by the pion chiral loop. The
moral here is that the contribution to $a_{\mu}(had\ 2)$ from chiral
loops, although $1/N_c$-suppressed, may well give rise to a sizeable
contribution.

There is yet another contribution to hadronic
light-by-light scattering which in $\chi$PT appears as a tree level
contribution to $O(p^6)$ in the chiral expansion; but contrary to the
$O(p^4)$ contribution which we have just discussed, it is leading in
the $1/N_c$ - expansion. It is the one induced by the Adler [22],
Bell and Jackiw [23] anomaly which in $\chi$PT is incorporated in
the so called Wess-Zumino-Witten term (WZW) [24,25]. When restricted
to two external photons and one pseudoscalar field, the WZW-term has
the simple form

$${\cal L}_{ABJ}={N_c\over 16\pi^2}{1\over
2}e^2\epsilon^{\mu\nu\rho\sigma}F_{\mu\nu}(x)F_{\rho\sigma}(x)
{1\over f_{\pi}}\left(\pi^0(x)+{1\over\sqrt 3}\eta (x)\right)\
.\eqno (35)$$
The $\pi^0(\eta)$ propagator between two effective
ABJ-vertices brings down the resulting $O(p^4)\times O(p^4)$
light-by-light scattering amplitude to $O(p^6)$; and because of the
factor $1/f_{\pi}^2$ the overall contribution is $O(N_c)$, and
therefore leading in the $1/N_c$-expansion. Obviously, this
contribution cannot be neglected; and in fact, the simple power
counting arguments above tell us that it could be the dominant
contribution.

The previous analysis of $a_{\mu} (had\ 2)$ within the framework of
$\chi$PT has served to identify three distinct sources of a priori
important contributions. We want now to discuss their possible
evaluation.

The contribution to $a_{\mu}(had\ 2)$ from the lowest order chiral
pion loop can be calculated unambiguously. It has in fact been
evaluated numerically by KNO using two different methods, with the
average result
$$a_{\mu}(had\ 2, \chi \hbox{ pion loop
})=-5.12(0.35)\times 10^{-10}\ .\eqno (36)$$
To this result one
should add the corresponding contribution from the chiral kaon loop;
presumably rather small.

A straightforward calculation of the contribution to $a_{\mu}(had\
2)$ from the $O(p^6)$ light-by-light scattering amplitude resulting
from the $\pi^0(\eta)$-exchange between two ABJ-vertices leads to
$UV$-divergent integrals. Once again, this is an indication that
more about the underlying QCD dynamics than just the local
$\pi^0(\eta)\gamma\gamma$ $O(p^4)$ coupling is needed. KNO, in their
evaluation of the ''$\pi^0$-exchange'' contribution, have solved the
problem by using a vector-meson-dominance (VMD) inspired
regularization of the three photon-propagators; i.e., they use the
prescription $Q^2_i\equiv -q^2_i\quad i=1,2,3)$ :
$${-1\over
Q^2_i}\to{-1\over Q^2_i}+{1\over Q^2_i+M^2_V}\ .\eqno (37)$$
It turns
out that this VMD - prescription is in fact very close to the result
one obtains in the ENJL-model [16], after summing all orders in
powers of momenta at the leading order in the $1/N_c$ - expansion,
which leads to the effective replacement :
$${-1\over Q^2_i}\to
{-1\over Q^2_i}{1\over 1+Q^2_i{8\pi^2 G_v\over N_c\Lambda^2_{\chi}}
\overline{\Pi}^{(1)}_V(Q^2_i)}={-M^2_V(Q^2_i)\over
Q^2_i\left[Q^2_i+M^2_V(Q^2_i)\right]}\ ,\eqno (38)$$
with
$\overline{\Pi}^{(1)}_V(Q^2)$ as given in eq.(20), and
$M_V(Q^2=0)=M_V$. Numerically, the result found by KNO from the
''$\pi^0$-exchange'' term, and using the VMD-prescription in (37),
is
$$a_{\mu}(\pi^0)=6.5(0.6)10^{-10}\ .\eqno (39)$$
The
corresponding contribution from $\eta$-exchange which should be
added to (39) is likely to be negligible.

An improved calculation of the ''$\pi^0$-exchange'' contribution
within the framework of the ENJL-model is possible. Schematically,
it corresponds to the replacement of the $\pi^0$-exchange  tree level
diagram in Fig. 3a by the one in Fig. 3b. The round dots $(\circ)$
in this figure represent four-fermion vector like couplings; the
square dots ($\carre$) four-fermion axial-vector like couplings. The
horizontal string of bubbles in Fig. 3b has already been calculated
(see section 3.2 of ref. [16]). It contains in particular the pion
(eta) - pole(s), but also axial-vector contributions. The three-point
functions at the vertices of Fig. 3b; i.e., the links
$\times\bigcirc^{\hbox{$\carre$}}_{\hbox{$\circ$}}$ and
$\carre\bigcirc^{\hbox{$\circ$}}_{\hbox{$\circ$}}$ , have
non-trivial $Q^2_i$ - dependence. In KNO's calculation, they have
been approximated by the corresponding local ABJ-vertices in (35).

Finally we turn our attention back to the leading $O(N_c)$
contribution we first discussed, which in $\chi$PT starts at
$O(p^8)$. This is an independent contribution which has to be
evaluated separately; and, contrary to what has been done so far in
the literature, it should be added to the two other contributions
previously discussed. KNO's numerical evaluation of what they call
the ''quark-loop'' contribution provides in my opinion a good first
approximation to the required calculation. The constituent
quark-loop graphs of refs. [3] and [26] correspond to the mean field
approximation of the ENJL-model of refs. [15,16]. In the case of
hadronic vacuum polarization, this is the approximation where
$\Pi^{(1)}_V\to\overline{\Pi}^{(1)}_V$ (see eqs. (19) and (20)). The
result found by KNO for the ''quark-loop'' contribution is
$$a_{\mu}(had\ 2, \hbox{ quark-loop })=6.0(0.4)\times 10^{-10}\ .
\eqno (40)$$
It corresponds to the choice $M_u=M_d=300 MeV$ and
$M_s=500 MeV$. (There is also a small contribution from the charm
quark with a mass value $M_c=1.5 GeV$ which has been included in
(40).)

Again, it is possible to do an improved calculation of the
''quark-loop'' type contribution within the framework of the
ENJL-model. The suggested improvement is schematically represented
in Fig. 4. In practice it amounts to the effective replacement
indicated in (38) for each photon-propagator. Since here we are
working in the chiral limit the $u,d$ and $s$ constituent quark
masses should be taken equal. I would expect a sizeable reduction of
the ''quark-loop'' estimate in (40) by perhaps as much as a factor
of two.

The main conclusion which emerges from our analyses is that the three
contributions in (36), (39), and (40) correspond to well identified
different sources in low-energy QCD. The three contributions should
be added in order to get an appropriate estimate of $a_{\mu}(had\
2)$. So far, only the result from the chiral $\pi$-loop contribution
in (36) can be considered to be fully reliable. (In that respect I
agree with the analyses of ref. [5].) Both the ''$\pi^0$-exchange''
and the ''quark-loop'' results in (39) and (40) should be considered
as first approximation estimates; specially the ''quark-loop''
result (40). These estimates can however be substantially refined if
one is ready to accept the ENJL-model as a good approximation to
low-energy QCD at large-$N_c$. We see no reason why the ENJL-model
should fail here, and the results we have obtained in the case of
the hadronic vacuum polarization are very encouraging. The
improvement can be done following the suggestions I have sketched
above. The implementation of this improvement is quite a formidable
task, but it should be relatively easy to incorporate in the already
existing computer programs developped by Kinoshita et al. A parallel
calculation of the \eject
\n $\gamma\gamma\to\gamma\gamma$
cross-section at low energies in the ENJL-model, and its comparison
to future experimental results of DA$\Phi$NE, could help to check
the validity of the model.

\bigskip

\n{\bf ACKNOWLEDGEMENTS}
\m
I am grateful to Hans Bijnens, Gilles Esposito-Far\`ese, Santi
Peris and Toni Pich for their comments and suggestions on the topics
discussed in this note.

\vfill\eject

\n {\bf REFERENCES :}
\m
\parindent 1truecm
\item{\hbox to\parindent{\enskip \hfill[1] \hfill}} J. Bailey {\it
et al.}, Phys. Lett., {\bf 68 B}, 191 (1977) ; F.J.M. Farley and E.
Picasso, {\it The muon $g-2$ Experiments}, Advanced Series on
Directions in High Energy Physics - Vol.~7 Quantum Electrodynamics,
ed. T. Kinoshita, World Scientific 1990.
\item{\hbox
to\parindent{\enskip \hfill[2] \hfill}} T. Kinoshita and W.J.
Marciano, {\it Theory of the Muon Anomalous Magnetic Moment},
Advanced Series on Directions in High Energy Physics - Vol. 7
Quantum Electrodynamics, ed. T. Kinoshita, World Scientific 1990.
\item{\hbox to\parindent{\enskip \hfill[3] \hfill}} T. Kinoshita, B
Ni\v zi\'c and Y. Okamoto, Phys. Rev., {\bf D 31}, (1985), 2108.
\item{\hbox to\parindent{\enskip \hfill[4] \hfill}} R. Barbieri and
E. Remiddi, {\it The Da$\phi$ne Physics Handbook} - Vol. II, p. 301 ;
ed. L.~Maiani {\it et al.}, Frascati, INFN 1992.
\item{\hbox
to\parindent{\enskip \hfill[5] \hfill}} M.B. Einhorn, {\it On the
Hadronic Contribution to Light-by-Light Scattering in $g-2$}~;
Preprint UM - TH - 93 - 18, (to be published in Phys. Rev.
Comments).
\item{\hbox to\parindent{\enskip \hfill[6] \hfill}} B.E.
Lautrup and E. de Rafael, Nuovo Cim. {\bf 64 A} (1969) 322~; B.E.
Lautrup, A.~Peterman and E. de Rafael, Phys. Rep. - Vol.~3~C,
N$^{\circ}$~4 (1972), 193-260.
\item{\hbox to\parindent{\enskip
\hfill[7] \hfill}} G. 't Hooft, Nucl. Phys., {\bf B 72}, (1974), 461.
\item{\hbox to\parindent{\enskip \hfill[8] \hfill}} J.S. Bell and E.
de Rafael, Nucl. Phys., {\bf B 11}, (1969), 611.
\item{\hbox
to\parindent{\enskip \hfill[9] \hfill}} J. Gasser and H. Leutwyler,
Ann. of Phys. (N.Y.) {\bf 158}, (1984) 142~; Nucl. Phys. {\bf B~250},
(1985), 465, 517, 539.
\item{\hbox to\parindent{\enskip [10] \hfill}}
G. Ecker, J. Gasser, A. Pich and E. de Rafael, Nucl. Phys. {\bf
B~321}, (1989), 311.
\item{\hbox to\parindent{\enskip [11] \hfill}}
J.F. Donoghue, C. Ramirez, and G. Valencia, Phys. Rev. {\bf D~39},
(1989), 1947.
\item{\hbox to\parindent{\enskip [12] \hfill}} G.
Ecker, J. Gasser, H. Leutyler, A. Pich, and E. de Rafael, Phys. Lett.
{\bf B~223}, (1989), 425.
\item{\hbox to\parindent{\enskip [13]
\hfill}} J.A. Casas, C. Lopez and F.J. Yndur\'ain, Phys. Rev. {\bf
D~32}, (1985), 736.
\item{\hbox to\parindent{\enskip [14] \hfill}}
L.M. Kurdadze {\it et al.}, Yad. Fiz. {\bf 40}, (1984), 451~; Sov. J.
Nucl. Phys. {\bf 40}, (1984), 286.
\item{\hbox to\parindent{\enskip
[15] \hfill}} J. Bijnens, Ch. Bruno and E. de Rafael, Nucl. Phys.
{\bf B~390}, (1993), 501.
\item{\hbox to\parindent{\enskip [16]
\hfill}} J. Bijnens, E. de Rafael and H. Zheng, {\it Low-Energy
Behaviour of Two-Point Func\-tions of Quark Currents}~; Preprint
CERN - TH. 6924~/~93, CPT - 93 / P.2917, Nordita - 93 / 43 N, P.
\item{\hbox to\parindent{\enskip [17] \hfill}} S. Peris and E. de
Rafael, Phys. Lett. {\bf B~309}, (1993), 389.
\item{\hbox
to\parindent{\enskip [18] \hfill}} A. Manohar and H. Georgi, Nucl.
Phys. {\bf B~234}, (1984), 1189.
\item{\hbox to\parindent{\enskip
[19] \hfill}} D. Espriu, E. de Rafael and J. Taron, Nucl. Phys. {\bf
B~345}, (1990), 22~; (Erratum {\bf B~355}, (1991), 278).
\item{\hbox to\parindent{\enskip [20] \hfill}} E. Euler, Ann. Phys.
(Leipzig) {\bf 26}, (1936), 398~; E. Euler and W. Heisenberg, Z.
Phys. {\bf 98}, (1936), 714.
\item{\hbox to\parindent{\enskip [21] \hfill}} J. Schwinger,
{\it Particles, Sources and Fields} - Vol II, Addision-Wesley Pub.
Company (1973).
\item{\hbox to\parindent{\enskip [22] \hfill}} S.L.
Adler, Phys. Rev. {\bf 177}, (1969), 2426.
\item{\hbox to\parindent{\enskip [23] \hfill}} J.S. Bell and R.
Jackiw, Nuovo Cim. {\bf 60~A}, (1969), 47.
\item{\hbox
to\parindent{\enskip [24] \hfill}} J. Wess and B. Zumino, Phys. Lett.
{\bf 37~B}, (1971), 95.
\item{\hbox to\parindent{\enskip [25]
\hfill}} E. Witten, Nucl. Phys. {\bf B~223}, (1983), 422.
\item{\hbox to\parindent{\enskip [26] \hfill}} J. Calmet,
P. Narison, M. Perrotet and E. de Rafael, Phys. Lett. {\bf
16~B}, (1976), 283~; Rev. Mod. Phys. {\bf 49}, (1977), 21.

\vfill\eject

\n {\bf FIGURE CAPTIONS :}
\m
\parindent 1,5truecm
\item{\hbox to\parindent{\enskip Fig. 1\hfill}} Hadronic vacuum
polarization to the second order muon $g-2$ (a)~; and to the
fourth-order muon $g-2$ (b).
\item{\hbox to\parindent{\enskip Fig.
2\hfill}} Hadronic light-by-light scattering contribution to the muon
$g-2$.
\item{\hbox to\parindent{\enskip Fig. 3\hfill}} Hadronic
light-by-light scattering contribution induced by the combination of
two three-point functions. Graph (a) represents the
$\pi^{\circ}$-exchange contribution. Graph (b) the corresponding
gene\-ra\-lized contribution within the ENJL-model. The square dots
($\carre$) represent four-fermion axial-vector like couplings~; the
round dots ($\circ$) four-fermion vector like couplings.
\item{\hbox
to\parindent{\enskip Fig. 4\hfill}} Hadronic light-by-light
scattering contribution induced by the irreducible four-point
function. Graph (a) represents the mean field approximation. Graph
(b) the full ENJL-model contribution to all orders in momenta for
the off-shell photons.

\vfill\eject





\end